# Superionic hydrogen in Earth's deep interior

Yu He[1,2], Duck Young Kim[1*], Chris J. Pickard[3,4], Richard J. Needs[5], Qingyang Hu[1],

Ho-kwang Mao[1]

[1] Centre for High Pressure Science and Technology Advanced Research (HPSTAR), Shanghai 201203, China

[2] Key Laboratory of High-Temperature and High-Pressure Study of the Earth's Interior, Institute of Geochemistry, Chinese Academy of Sciences, Guiyang, Guizhou 550081, China

[3] Department of Materials Science and Metallurgy, University of Cambridge, 27 Charles Babbage Road, Cambridge Cb3 0Fs, United Kingdom

[4] Advanced Institute for Materials Research, Tohoku University 2-1-1 Katahira, Aoba, Sendai, 980-8577, Japan

[5] Theory of Condensed Matter Group, Cavendish Lavoratory, JJ Thomson Avenue, Cambridge CB3 0HE, United Kingdom

* Corresponding author

*Email address: duckyoung.kim@hpstar.ac.cn (D. Y. Kim)*

**Superionic hydrogen was previously thought to be an exotic state predicted and confirmed only in pure $H_2O$ ice[1-3]. In Earth's deep interior, $H_2O$ exists in the form of O-H groups in ultra-dense hydrous minerals, which have been proved to be stable even at the conditions of the core-mantle boundary (CMB)[4-11]. However, the superionic states of these hydrous minerals at high *P-T* have not been investigated. Using first-principles calculations, we found that pyrite structured $FeO_2H_x$ ($0 \leqslant x \leqslant 1$) and δ-AlOOH, which have been proposed to be major hydrogen-bearing phases in the deep lower mantle (DLM), contain superionic hydrogen at high *P-T* conditions. Our observations indicate a universal pathway of the hydroxyl O-H at low pressure transforming to symmetrical O-H-O bonding at high-*P* low-*T*, and a superionic state at high-*P* high-*T*. The superionicity of hydrous minerals has a major impact on the electrical conductivity and hydrogen transportation behaviors of Earth's lower mantle as well as the CMB.**

Hydrogen, the most abundant element in the solar system, plays a pivotal role in Earth[12-15]. It forms water that covers 70% of Earth's surface, it is the unique biochemical ingredient that makes Earth a living planet, it has drastic effects on physical and chemical properties of minerals and rocks, and it is the dominant component of volatiles that dictate Earth's evolution processes. Our knowledge on the nature of hydrogen in Earth's deep interior is quite limited and depends upon studies of mineral physics[4-11,16,17]. Very recent discoveries of a series of high-pressure ultradense hydrous phases, including the δ-AlOOH[4,5] and phase H[6] hydrous aluminum magnesian silicate[7,8], pyrite

$FeO_2H$[9,10], and $(Al,Fe)O_2H$[11] suggest the presence of a significant amount of hydrogen in the DLM. The present study further reveals that the hydrogen under the DLM pressure-temperature (*P-T*) conditions is present in the superionic state that has entirely different properties from the ordinary O-H bonding, and has profound impacts on geophysical and geochemical behavior of Earth.

In the superionic state the hydrogen ion is no longer bonded to an oxygen ion, and it moves freely within the crystal lattice, leading to greatly improved ionic and electronic transport properties and effortless removal of hydrogen. Superionic hydrogen was proposed and observed in $H_2O$ ice[1-3]. However, hydrogen will not be present in the DLM in the form of free $H_2O$, and instead it will exist as new ultradense hydrous phases. Here we investigate two archetypal phases, pyrite $FeO_2H$ and δ-AlOOH, and discover superionic hydrogen to be a common behavior of hydrogen-bearing compounds, thus opening a new paradigm in DLM.

Fe-O binary compounds have previously been investigated using crystal structure searches combined with first-principles calculations, and the cubic $FeO_2$ structure has been identified as a stable phase[18,19]. We have performed crystal structure searches using the Ab Initio Random Structure Searching (AIRSS) method to study ternary materials consisting of Fe, O, and H at 100 GPa. The results of searches over 495 possible compositions are shown in Extended Data Fig. 1, and in total we have relaxed ~53,000 structures to minima in the energy landscape. In Fig. 1 we display a schematic summary of stable compounds at 100 GPa shown as red circles. Including previously known compounds, we also found pyrite-structured cubic py-$FeO_2$ and py-$FeO_2H$, which are consistent with experimental observations[9,10,19,20].

Furthermore, along the $FeO_2$ and $FeO_2H$ convex hull line, we found that $FeO_2H_x$ can be stabilized as either a ground state ($FeO_2H_{0.75}$) or a meta-stable state ($FeO_2H_{0.5}$ and $FeO_2H_{0.25}$). The inset in Fig. 1 shows the formation enthalpy along the line. Having $FeO_2$ and $FeO_2H$ as the end members, $FeO_2H_{0.75}$ becomes the ground state and $FeO_2H_{0.5}$ and $FeO_2H_{0.25}$ are close to the stability line. $FeO_2H_{0.75}$ possesses a rhombohedral structure, which is slightly distorted from the cubic pyrite structure with an angle of ~0.4 degrees. The small angular distortion arises from the hydrogen defects which are difficult to observe in experiments[21]. The dynamic stability of $FeO_2H_{0.75}$ is also verified by the stability of the phonons calculated at 100 GPa (Extended Data Fig. 1c). The partial release of hydrogen from $FeO_2H$ is further discussed in Methods.

Hydrogen in $FeO_2H$ possesses symmetric O-H-O bond, which are equivalent to those in ice X[22,23]. To further investigate the hydrogen bonding in iron hydrate we calculated the O-H bond length ($R_{O-H}$) as a function of the nearest-neighbor O-O bond length ($R_{O-O}$) for α-FeOOH (Goethite), ε-FeOOH, and py-$FeO_2H$. The bond lengths are compared in Fig. 2. In α-FeOOH (<10 GPa), hydrogen is closely bonded to a single O atom. The bond length of O-H increases slightly under compression. Delocalization of the H atom between the two O atoms can readily be seen in the ε phase[24]. The H atoms move to the center of the two O atoms with $R_{O-H} = R_{O-O}/2$, when the O-O distance is reduced to 2.35 Å. The symmetric O-H-O bond is retained in py-FeO$_2$H. The symmetrization of the hydrogen bond has also been reported in other hydrous phases, such as δ-AlOOH[5] and phase D (MgSi$_2$O$_6$H$_2$)[25] at high pressure. The evolution of O-H-O bonds and the localized charge density can be better visualized using the electron localization function (ELF) shown in Fig. 2. In its asymmetric states H is covalently bonded with one oxygen forming hydroxyl (OH$^-$). During the H centering process high pressure hydrogen becomes ionic (H$^+$) in the crystalline form. The covalent-to-ionic hydrogen transition in iron hydrate and δ-AlOOH at high-pressure is quite similar to water (gray stars and diamonds in Fig. 2)[22,23]. It is known that under high pressure water possesses a superionic state at pressures above 50 GPa and temperatures above 2000~K, with potential influence on the interiors of Neptune and Uranus[2,3]. Considering the similar O-H-O bonds evolution in the ultradense hydrous minerals, they may also possess a superionic state with implications for Earth's deep lower mantle.

AIMD calculations were used to simulate proton diffusion in py-FeO$_2$H$_x$ (x=1 and 0.5) and δ-AlOOH at temperatures and pressures within the range 1000-3500~K and 60-140 GPa. Simulations of up to about 7 ps were run for py-FeO$_2$H$_x$ and 14 ps for δ-AlOOH. At low temperatures protons in the hydrous phases undergo small mean-square displacements (MSDs) which do not increase with time and simply vibrate about their lattice positions, demonstrating that the material is an ordinary solid. Increasing the temperature leads to highly diffusive protons with the MSD increasing monotonically during the simulation, while the O, Fe and Al ions vibrate harmonically and adopt centered lattice positions, indicating a superionic state. A further increase in temperature leads to melting of the lattice and the ions diffuse away from their equilibrium positions and their MSDs increase with simulation time (Extended Data Fig. 3 and Extended Data Fig. 4). The calculated phase diagrams of py-FeO$_2$H, py-FeO$_2$H$_{0.5}$ and δ-AlOOH showing their solid-superionic-melting regions are plotted in Fig. 3. The hydrous phases can become superionic as the temperature increases. The superionic temperature range of py-FeO$_2$H is around 2250~K at pressures below 100 GPa and 2500-2750~K at pressures above 100~GPa (green region). This superionic region covers part of the geotherm curve in the lower mantle[26]. On the other hand, the superionic temperature range of py-FeO$_2$H$_{0.5}$ is much wider, being from 1750 to 2500~K. The relatively low solid-superionic phase transition temperatures of py-FeO$_2$H$_{0.5}$ arise from the existence of hydrogen vacancies, which provide more sites for proton migration. As the hydrogen concentration in py-FeO$_2$H$_x$ is normally around 50-70%, py-FeO$_2$H$_x$ can become superionic at lower mantle conditions. δ-AlOOH transforms to its superionic state at the temperatures above 2750~K. This transition temperature is above the geotherm of the lower mantle. However, the temperature increases rapidly to over 3000~K at the CMB and hydrogen in δ-AlOOH can also be superionic under these conditions. We did not observed the melting of the δ-AlOOH lattice at the pressure and temperature ranges of our simulations.

In order to calculate diffusion coefficients and ionic conductivity of py-FeO$_2$H$_x$ (x=1

and 0.5) we performed AIMD calculations with long simulation runs (15 ps) at pressures around 80 and 130 GPa. Using the MSDs (Extended Data Fig. 5), we calculated proton diffusion coefficients at different temperatures, which were fitted to an Arrhenius equation as shown in Extended Data Fig. 6. The deduced activation enthalpy of py-FeO$_2$H$_{0.5}$ is much lower than that of py-FeO$_2$H. We calculated the ionic conductivity using the Nernst-Einstein equation as plotted in Fig. 4. The separated symbols give the calculated proton conductivities of py-FeO$_2$H$_x$ deduced from their MSDs, and the dashed lines in Fig. 4 show the conductivity of py-FeO$_2$H$_x$ deduced from the linear fit to the diffusion coefficients. The conductivities of py-FeO$_2$H$_x$ at the superionic state with short simulation runs (7 ps) are also labeled by small stars in Fig. 4. This reveals that the proton conductivities in py-FeO$_2$H$_x$ are distributed between 5 S m$^{-1}$ to 600 S m$^{-1}$, varying with applied temperature, pressure and the concentration of hydrogen vacancies. Generally, py-FeO$_2$H$_{0.5}$ with 50% hydrogen vacancies shows higher ionic conductivity at relatively lower temperatures (< 2300~K). However, the conductivity of py-FeO$_2$H without hydrogen vacancies increases rapidly with temperature, which leads to a higher conductivity at temperatures above 2300~K. The difference between the conductivities of py-FeO$_2$H$_{0.5}$ at 80 and 130 GPa decrease with increasing temperature, and the maximum difference is less than an order of magnitude over the temperature range in which the superionic state occurs. The conductivity of py-FeO$_2$H at 80 GPa is about an order of magnitude higher than that at 130 GPa within the narrow superionic temperature range (2500-2750~K). The reduction in the diffusion rate of the protons at high pressure is attributed to the increase in the proton migration barrier energy with pressure (Extended Data Fig. 7). The calculated proton conductivities of δ-AlOOH are also exhibited in Fig. 4. The highest proton conductivity of δ-AlOOH reaches 100 S m$^{-1}$ when temperature is over 3500~K. The conductivities of the two hydrous phases are compared with the measured conductivities of Fe-bearing Bridgmanite and the post-perovskite phase (Fig. 4)[27-29]. The ionic conductivities of py-FeO$_2$H$_x$ and δ-AlOOH are mostly higher than the conductivity of Bridgmanite. At the conditions of the CMB with temperature higher than 2500~K, the ionic conductivities of py-FeO$_2$H$_x$ are comparable to that of the post-perovskite phase (>100 S m$^{-1}$), while the conductivities of δ-AlOOH can exceed 100 S m$^{-1}$ only at the temperature above 3500~K.

From the point of view of geoscience, with the essentially unlimited iron reservoir in the core and the water brought by subduction from the crust and the primordial water in the mantle[30], significant amount of py-FeO$_2$H$_x$ can be deposited at the CMB[20]. During the formation of py-FeO$_2$H$_x$, certain amount of hydrogen vacancies are generated in the lattice[10,31]. These vacancies provide diffusion sites for protons and lead to the superionic state of py-FeO$_2$H$_x$ at the conditions of the lower mantle and the CMB (Fig. 3). δ-AlOOH with a wide stable pressure and temperature range can be brought down to the CMB by subduction process and mantle convection. At the CMB with temperature exceeding 3000~K, hydrogen in δ-AlOOH can be superionic. Beyond py-FeO$_2$H$_x$ and δ-AlOOH, other hydrous minerals with symmetric ionic hydrogen bonding may also transform to superionic state at high temperature. In the superionic state of

py-FeO$_2$H$_x$, δ-AlOOH, and/or other hydrous minerals, proton diffusion in the superionic state generates an electric current at the bottom of the mantle which leads to electromagnetic interactions at the boundary of the geodynamo. This changes the boundary conditions of the geodynamo and affects the geomagnetic field. This results may offer a clue on the geodynamo simulations to understand some geomagnetic field characteristics, which have not been well understood, such as the slight ripples in the field intensity and the reversals of Earth's magnetic field. In addition, the ionic conductivities of py-FeO$_2$H$_x$ and δ-AlOOH reach 100 S m$^{-1}$ at the CMB, which is comparable with the conductivity of the post-perovskite phase (Fig. 4). The hydrous minerals along with the post-perovskite phase can then form the high conductivity layer (> 10$^8$ S) at CMB, which enhances the electromagnetic (EM) coupling between the fluid core and solid mantle and results in a change in the length of a day[27].

Water from different origins has different abundances of isotopes for both H and O. When py-FeO$_2$H$_x$, δ-AlOOH, and/or other hydrous minerals transform to their superionic states, H$^+$ and D$^+$ become highly diffusive like liquid while other ions (O, Fe, and Al et al.) remain close to their initial positions. H$^+$ and D$^+$ can be fully mixed during this procedure, but O isotopic compositions will derive from the same origins. As a result of this process, H and O circulations in Earth are separated, and hydrogen reservoirs with uniform isotopic composition form in the lower mantle and CMB. The protons released from the hydrous minerals incorporate other mantle minerals which can be transported back to Earth's surface and complete the hydrogen circulation.

Coincidentally, eliminating the contribution of sea water, the H isotopic composition is homogeneous, in contrast to the isotopic composition of elemental O, in mid-ocean ridge basalts (MORB) with a D value of -80±10[32] or -60±5[33]. The existence of the superionic hydrogen reservoirs in the lower mantle can provide sufficient uniform isotopic hydrogen for MORB.

**Methods**

**Ab Initio Random Structure Searching (AIRSS)**

Density functional theory (DFT) calculations were performed using the generalized gradient approximation (GGA) functionals, in particular the Perdew-Burke-Ernzerhof (PBE) exchange-coorelation functional[34]. Crystal structure searching was conducted using Ab Initio Random Structure Searching (AIRSS, version 0.91)[35,36] combined with planewave-based DFT softwareusing the CASTEP code[37]. A plane-wave basis set cutoff energy of 280 eV and k-point sampling of 0.06 Å$^{-1}$ were used.

**Structure relaxation and molecular dynamics simulations**

DFT calculations for the structure relaxation and molecular dynamics simulations were performed using the projector augmented wave (PAW) method[38] as implemented in the Vienna Ab Inito Simulaton Package (VASP) package[39]. In our calculations, a plane wave representation for the wave function with a cutoff energy of 800 eV was adopted. Geometry optimizations were performed using conjugate gradients minimization until all of the forces acting on the ions were less than 0.01 eV/Å per atom. K-point mesh with a spacing of ca. 0.03 Å$^{-1}$ was adopted. Considering the strongly correlated nature of the Fe 3d electrons a Hubbard-type U correction was used[40,41]. In accord with previous work, the effective U value was set to 5 eV[42]. H vacancies in py-FeO$_2$H$_x$ (x = 0.75, 0.5 and 0.25) were generated by removing H atoms from a large supercell of py-FeO$_2$H comprising 2 × 2 × 2 conventional unit cells, as shown in the Extended Data Fig. 8.

**Calculations of the equation of state (EOS)**

Structural relaxations were performed at various constant volumes and the calculated energy-volume data at zero Kelvin was fitted to a third-order Birch-Murnaghan equation of state (EOS):

$$E(V) = E_0 + \frac{9V_0 B_0}{16}\left\{\left[\left(\frac{V_0}{V}\right)^{\frac{2}{3}} - 1\right]^3 B_0' + \left[\left(\frac{V_0}{V}\right)^{\frac{2}{3}} - 1\right]^2 \left[6 - 4\left(\frac{V_0}{V}\right)^{\frac{2}{3}}\right]\right\} (1),$$

where $E_0$ denotes the intrinsic energy at zero pressure, $V_0$ is the volume at zero pressure, $B_0$ is the bulk modulus, and $B_0'$ is the first pressure derivative of the bulk modulus. The fitted parameters at zero Kelvin are shown in Extended Data Table 1. The relationship between the pressure and volume at zero Kelvin can be expressed as:

$$P(V) = \frac{3B_0}{2}\left[\left(\frac{V_0}{V}\right)^{\frac{7}{3}} - \left(\frac{V_0}{V}\right)^{\frac{5}{3}}\right]\left\{1 + \frac{3}{4}(B_0' - 4)\left[\left(\frac{V_0}{V}\right)^{\frac{2}{3}} - 1\right]\right\} \quad (2).$$

The pressure-volume relations at higher temperatures are plotted in Extended Data Fig. 9. The P-V relation of py-FeO$_2$H fits the experimental data very well, while the calculated volume of py-FeO$_2$ is a little overestimated compared with experimental data[9,10,19]. It is generally accepted that DFT calculations with a GGA-PBE density functional usually overestimate cell volumes. In this case the cell volumes of py-FeO$_2$H$_x$ decrease with hydrogen concentration. Our calculations also show that the hydrogen concentrations of py-FeO$_2$H$_x$ samples reported by Hu *et al.* are mostly above 50%[10]. The phase stability of py-FeO$_2$H$_x$ was investigated by calculating relative enthalpies with respect to py-FeO$_2$H. As shown in Extended Data Fig. 9b, the relative enthalpies of the solid solutions of the py-FeO$_2$H$_x$ phases (x = 0.75, 0.5 and 0.25) (solid lines) are lower than that of the separate xFeO$_2$H + (1-x)FeO$_2$ phases (dashed line) in the pressure range 50-210 GPa. This means that FeO$_2$ and FeO$_2$H can react with each other to form FeO$_2$H$_x$, which is energetically stable (Eq. 3). This is consistent with previous experimental and theoretical investigations[10,31,43,44].

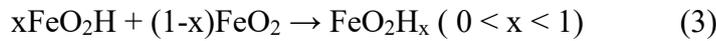

$$xFeO_2H + (1-x)FeO_2 \rightarrow FeO_2H_x \quad (0 < x < 1) \quad (3)$$

**Calculations of proton migration path and barrier energy**

Calculations of the proton migration path and barrier activation energy for a proton in a py-FeO$_2$H$_x$ lattice at different pressures were obtained using the Climbing-image nudged elastic band (CINEB) method[45] in a 2 × 2 × 2 supercell containing 128 atoms.

With fixed starting and end points, this approach duplicated a series of images (seven in our calculations) between the starting and end point of the migration of the ion so that the intermediate states could be simulated. For the large supercell adopted in the CINEB calculations, which tests showed to be sufficient for drawing qualitatively correct conclusions.

The calculated proton migration paths and a barrier energy in py-FeO$_2$H$_x$ are illustrated in Extended Data Fig. 7. Only one hydrogen vacancy was generated at a time in a 2 × 2 × 2 supercell of py-FeO$_2$H. Due to the cubic symmetry of the system the possible migration paths are along the <110>, <101> and <011> directions (Extended Data Fig. 7b). As a result, py-FeO$_2$H$_x$ provides a three-dimensional (3D) path for proton diffusion. The corresponding barrier energies increase slightly with pressure from 1.88 to 2.01 eV (Extended Data Fig. 7c).

**Ab initio molecular dynamics (AIMD) calculations of proton diffusion**

To calculate the H$^+$ diffusion coefficient we performed AIMD simulations[1-3,46-50] using a 2 × 2 × 2 supercell (128 atoms) for FeO$_2$H$_x$ and a 2 × 2 × 4 supercell (128 atoms) for δ-AlOOH. It is worth to note that a bigger supercell does not change the simulation results. The simulations used the canonical ensemble with a time step of 1 fs, with the simulations lasting 7ps and 15 ps at temperatures from 1250~K to 3500~K. This approach is suitable for evaluating the activation enthalpy of ion migration and identifying the ion transport mechanism. To study superionic transport rigorously, we calculated the diffusion coefficient for $H^+$ transport and the mean-square displacement (MSD) of the ionic positions. The diffusion coefficient is defined as:

$$D = \lim_{0 \to \infty} \left[ \frac{1}{2dt} \langle [\vec{r}(t)]^2 \rangle \right] \quad (4),$$

where $d$ is the dimension of the lattice on which ion hopping takes place. The MSD

$$\langle [\vec{r}(t)]^2 \rangle = \frac{1}{N} \sum_{i=1}^{N} \langle [\vec{r_i}(t+t_0) - \vec{r_i}(t_0)]^2 \rangle \quad (5),$$

is averaged over all protons, and $\vec{r_i}(t)$ is the displacement of the ith proton at time $t$, and $N$ is the total number of protons in the system. In practice, $D$ is obtained by a linear fit to the time dependence of the average MSD.

The value of D obtained at various temperatures can be fitted with an Arrhenius equation:

$$D = A \exp\left(-\frac{\Delta H}{kT}\right) \quad (6),$$

where $\Delta H$ is the activation enthalpy, $A$ is a pre-exponential factor, $k$ is the Boltzmann constant, and $T$ is the temperature. The calculated proton diffusion coefficients in py-$FeO_2H_x$ and $\delta$-$AlOOH$ are plotted in Extended Data Fig. 10. The electrical conductivity was calculated using the diffusion coefficients and the Nernst-Einstein equation:

$$\sigma = \frac{fDcq^2}{kT} \quad (7),$$

in which $\sigma$ is the electrical conductivity, $f$ is a numerical factor approximately equal to unity, $D$ is the diffusion coefficient, $c$ is the concentration of $H^+$ vacancies, $q$ is the electrical charge of $H^+$, $k$ is the Boltzmann constant, and $T$ is the temperature. The superionic state of py-FeO$_2$H$_x$ is also seen clearly in the 15 ps simulations and in the MSDs and the trajectories of the protons (Extended Data Fig. 5 and Extended Data Fig. 11). At higher temperatures, crystalline py-FeO$_2$H$_x$ becomes unstable and starts to melt while the MSDs of the Fe and O ions increase with simulation time (Extended Data Fig. 12). The experiments show that py-FeO$_2$H$_x$ is stable even at 3100-3300~K and pressures of 126 GPa[43]. This discrepancy is within the range expected from data for pure iron[51] and iron oxides[52]. We note here that a sophisticated theory employed to calculate melting temperatures might improve the estimation[51,53]. For δ-AlOOH, it shows the anisotropic proton diffusion with a higher proton MSD along the [001] direction as shown in Extended Data Fig. 13. This is due to the channel-like structure of δ-AlOOH along [001] direction.

**Reference:**

1. Demontis, P., LeSar, R. & Klein, M. L. New high-pressure phases of ice. Phys. Rev. Lett. **60**, 2284-2287 (1988).

2. Cavazzoni, C. et al. Superionic and metallic states of water and ammonia at giant planet conditions. Science **283**, 44-46, (1999).

3. Millot, M. et al. Experimental evidence for superionic water ice using shock compression. Nature Phys. **14**, 297–302 (2018).

4. Sano, A. et al. Aluminous hydrous mineral d-AlOOH as a carrier of hydrogen into the core-mantle boundary. Geophys. Res. Lett. **35**, L03303 (2008).

5. Panero, W. R. & Stixrude, L. P. Hydrogen incorporation in stishovite at high

pressure and symmetric hydrogen bonding in δ-AlOOH. Earth Planet. Sci. Lett. **221**, 421-431 (2004).

6. Nishi, M. et al. Stability of hydrous silicate at high pressures and water transport to the deep lower mantle. Nature Geosci. **7**, 224-227 (2014).

7. Ohira, I. et al. Stability of a hydrous δ-phase, AlOOH-MgSiO$_2$(OH)$_2$, and a mechanism for water transport into the base of lower mantle. Earth Planet. Sci. Lett. **401**, 12-17 (2014).

8. Pamato, M. G. et al. Lower-mantle water reservoir implied by the extreme stability of a hydrous aluminosilicate. Nature Geosci. **8**, 75-79 (2015).

9. Nishi, M., Kuwayama, Y., Tsuchiya, J. & Tsuchiya, T. The pyrite-type high-pressure form of FeOOH. Nature **547**, 205-208 (2017).

10. Hu, Q. et al. Dehydrogenation of goethite in Earth's deep lower mantle. Proc. Natl Acad. Sci. USA **114**, 1498–1501 (2017).

11. Zhang, L., Yuan, H., Meng, Y. & Mao, H.-k. Discovery of a hexagonal ultradense hydrous phase in (Fe,Al)OOH. Proc. Natl. Acad. Sci. U. S. A. **115**, 2908-2911 (2018).

12. Williams, Q. & Hemley, R. J. Hydrogen in the deep earth. Annu. Rev. Earth Planet. Sci. **29**, 365-418 (2001).

13. Jacobsen, S. D. & van der Lee, S. Earth's deep water cycle. Vol. 168 (American Geophysical Union, 2006).

14. Bell, D. R. & Rossman, G. R., Water in Earth's Mantle: The Role of Nominally Anhydrous Minerals. Science **255**, 1391-1397 (1992).

15. Murakami, M., Hirose, K., Yurimoto, H., Nakashima, S. & Takafuji, N. Water in Earth's lower mantle. Science **295**, 1885-1887 (2002).

16. Matsukage, K. N., Jing, Z. C. & Karato, S. Density of hydrous silicate melt at the conditions of Earth's deep upper mantle. Nature **438**, 488-491 (2005).

17. Wang, D. J., Mookherjee, M., Xu, Y. S. & Karato, S. The effect of water on the electrical conductivity of olivine. Nature **443**, 977-980 (2006).

18. Weerasinghe G. L., Pickard, C. J. & Needs, R. J. Computational searches for iron oxides at high pressures. J. Phys.: Condens. Matter 27, 455501 (2015).

19. Hu, Q. et al. $FeO_2$ and FeOOH under deep lower-mantle conditions and Earth's oxygen–hydrogen cycles. Nature 534, 241–244 (2016).

20. Mao, H. K. et al. When water meets iron at Earth's core-mantle boundary. Natl. Sci. Rev. 4, 870-878 (2017).

21. Mao, W. L. et al. Distortions and stabilization of simple-cubic calcium at high pressure and low temperature, Proc. Natl Acad. Sci. USA **107**, 9965-9968 (2010).

22. Goncharov, A. F., Struzhkin, V. V., Somayazulu, M. S., Hemley, R. J. & Mao, H. K. Compression of ice to 210 gigapascals: Infrared evidence for a symmetric hydrogen-bonded phase. Science **273**, 218-220 (1996).

23. Benoit, M., Marx, D. & Parrinello, M. Tunnelling and zero-point motion in high-pressure ice. Nature **392**, 258-261 (1998).

24. Gleason, A. E., Quiroga, C. E., Suzuki, A., Pentcheva, R. & Mao, W. L. Symmetrization driven spin transition in ε-FeOOH at high pressure. Earth Planet. Sci. Lett. **379**, 49-55 (2013).

25. Hushur, A., Manghnani, M. H., Smyth, J. R., Williams, Q., Hellebrand, E., Lonappan, D., Ye, Y., Dera, P. & Frost, D. J. Hydrogen bond symmetrization and equation of state of phase D. J. Geophys. Res. 116, B06203 (2011).

26. Brown, J. M. & Shankland, T. J. Thermodynamic parameters in the Earth as determined from seismic profiles. Geophys. J. R. Astron. Soc. 66, 579–596 (1981).

27. Ohta, K. et al. The electrical conductivity of post-perovskite in Earth's D" layer. Science **320**, 89–91 (2008).

28. Katsura, T., Sato, K. & Ito, E. Electrical conductivity of silicate perovskite at lower-mantle conditions. Nature **395**, 493–495 (1998).

29. Yoshino, T., Yamazaki, D., Ito, E. & Katsura, T. No interconnection of ferro-periclase in post-spinel phase inferred from conductivity measurement. Geophys. Res. Lett. 35, L22303 (2008).

30. Hallis, L. J. et al. Evidence for primordial water in Earth's deep mantle. Science **350**, 795-797 (2015).

31. Yuan, L. et al. Chemical reactions between Fe and $H_2O$ up to megabar pressures and implications for water storage in the earth's mantle and core. Geophys. Res. Lett. **45**, 1330–1338 (2018).

32. Kyser, T. K. & O'Neil, J. R. Hydrogen isotope systematics of submarine basalts. Geochim. Cosmochim. Acta 48, 2123-2133 (1984).

33. Clog, M., Aubaud, C., Cartigny, P. & Dosso, L. The hydrogen isotopic composition and water content of southern Pacific MORB: A reassessment of the D/H ratio of the depleted mantle reservoir. Earth Planet. Sci. Lett. 381, 156-165 (2013).

34. Perdew, J. P., Burke, K. & Ernzerhof, M. Generalized gradient approximation made simple. *Phys. Rev. Lett.* **77,** 3865–3868 (1996).

35. Pickard, C. J. & Needs R. J. Ab initio random structure searching. J. Phys.: Condens. Matter 23, 053201 (2011).

36. Sun, J., Martinez-Canales, M., Klug, D. D., Pickard, C. J. & Needs, R. J. Phys. Rev. Lett. Persistence and eventual demise of oxygen molecules at terapascal pressures. 108, 045503 (2012).

37. Clark, S. et al. First principles methods using CASTEP: Zeitschrift für Kristallographie - Crystalline Materials. Z. Kristallogr. 220, 567 (2005)

38. Blöchl, P. E. Projector augmented-wave method. Phys. Rev. B 50, 17953–17979 (1994).

39. Kresse, G. Efficient iterative schemes for *ab initio* total-energy calculations using a plane-wave basis set. Phys. Rev. B 54, 11169–11186 (1996).

40. Dudarev, S. L., Botton, G. A., Savrasov, S. Y., Humphreys, C. J. & Sutton, A. P. Electron-energy loss spectra and the structural stability of nickel oxide: An LSDA+U study. Phys. Rev. B 57, 1505 (1998).

41. Liechtenstein, A. I., Anisimov, V. I. & Zaanen, J. Density-functional theory and strong interactions: Orbital ordering in Mott-Hubbard insulators. Phys. Rev. B 52,

R5467 (1995).

42. Jang, B. G., Kim, D. Y. & Shim, J. H. Metal-insulator transition and the role of electron correlation in $FeO_2$. Phys. Rev. B 95, 075144 (2017).

43. Liu, J. et al. Hydrogen-bearing iron peroxide and the origin of ultralow-velocity zones. Nature 551, 494-497 (2017).

44. Zhu, S. C., Hu, Q., Mao, W. L., Mao, H. K. & Sheng, H. Hydrogen-bond symmetrization breakdown and dehydrogenation mechanism of $FeO_2H$ at high pressure. J. Am. Chem. Soc. 139, 12129-12132 (2017).

45. Henkelman, G., Uberuaga, B. P. & Jonsson, H. A climbing image nudged elastic band method for finding saddle points and minimum energy paths. J. Chem. Phys. 113, 9901–9904 (2000).

46. Mookherjee, M., Stixrude, L. & Karki, B. Hydrous silicate melt at high pressure. Nature **452**, 983-986 (2008).

47. Wilson, H. F., Wong, M. L. & Militzer, B. Superionic to superionic phase change in water: Consequences for the interiors of Uranus and Neptune. Phys. Rev. Lett. **110**, 151102 (2013).

48. Hernandez, J. A. & Caracas, R. Superionic-Superionic Phase transitions in body-centered cubic $H_2O$ ice. Phys. Rev. Lett. **117**, 135503 (2016).

49. He, Y. et al. First-principles prediction of fast migration channels of potassium ions in $KAlSi_3O_8$ hollandite: Implications for high conductivity anomalies in subduction zones. Geophys. Res. Lett. **43**, 6228-6233 (2016).

50. Sun, J., Clark, B. K., Torquato, S. & Car R. The phase diagram of high-pressure

superionic ice. Nat. Commun. **6**, 8156 (2015).

51. Alfè, D. Temperature of the inner-core boundary of the Earth: Melting of iron at high pressure from first-principles coexistence simulations. Phys. Rev. B 79 060101 (2009).

52. Zhang, L. & Fei, Y. Melting behavior of (Mg,Fe)O solid solutions at high pressure. Geophys. Res. Lett. 35, L13302 (2008).

53. Belonoshko, A. B., Ahuja, R. & Johansson, B. Quasi-Ab Initio Molecular Dynamics Study of Fe Melting. Phys. Rev. Lett. 84, 3638-3641 (2000).

**Acknowledgements**

We acknowledge the support of National Natural Science Foundation of China (Grant No. 41774101). C.J.P. is supported by the Royal Society through a Royal Society Wolfson Research Merit award and the EPSRC through Grants No. EP/P022596/1. We are grateful to Dr. Y. Sun for MD technical supports and discussion.

**Author Contributions**

Y. H, D. Y. K, C. J. P, Q. H. conducted computational study and all authors analyze the data and wrote the manuscript.

**Author Information**

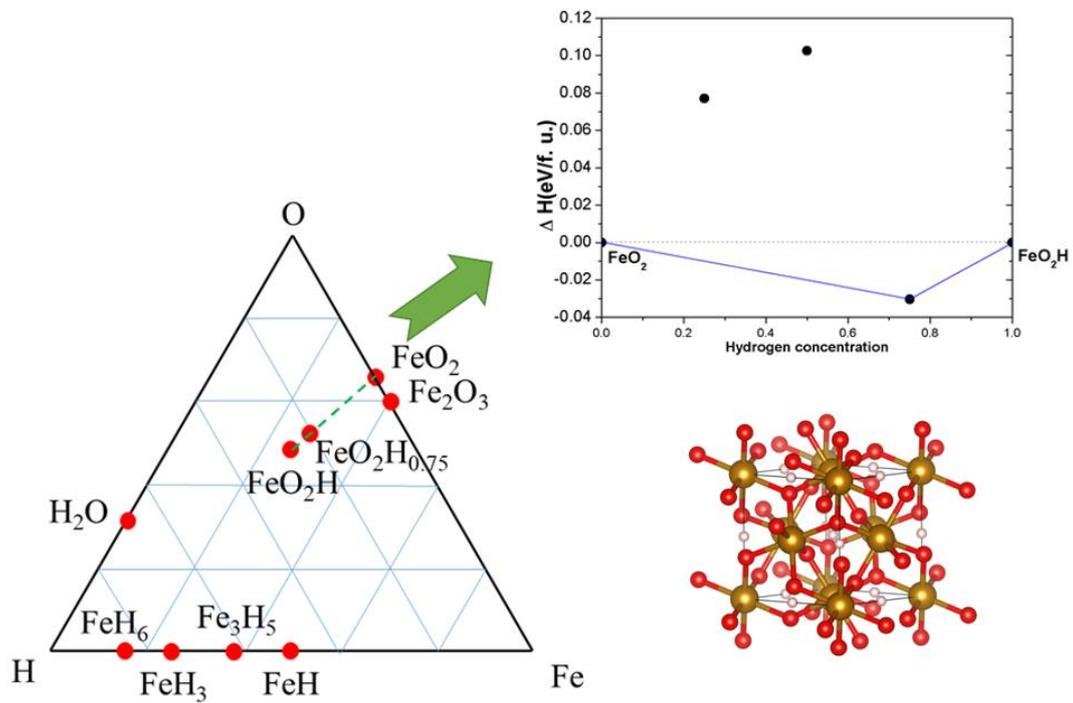

**Figure 1 | Structure searching for ternary materials consisting of Fe, O, and H at 100 GPa.** In the Fe, O and H phase diagram, the solid red circles are stable compounds. The inset shows a partial convex hull along the green dashed line from py-FeO$_2$ to py-FeO$_2$H. The crystal structure of py-FeO$_2$H is also shown with brown, red and pink spheres representing the Fe, O and H atoms.

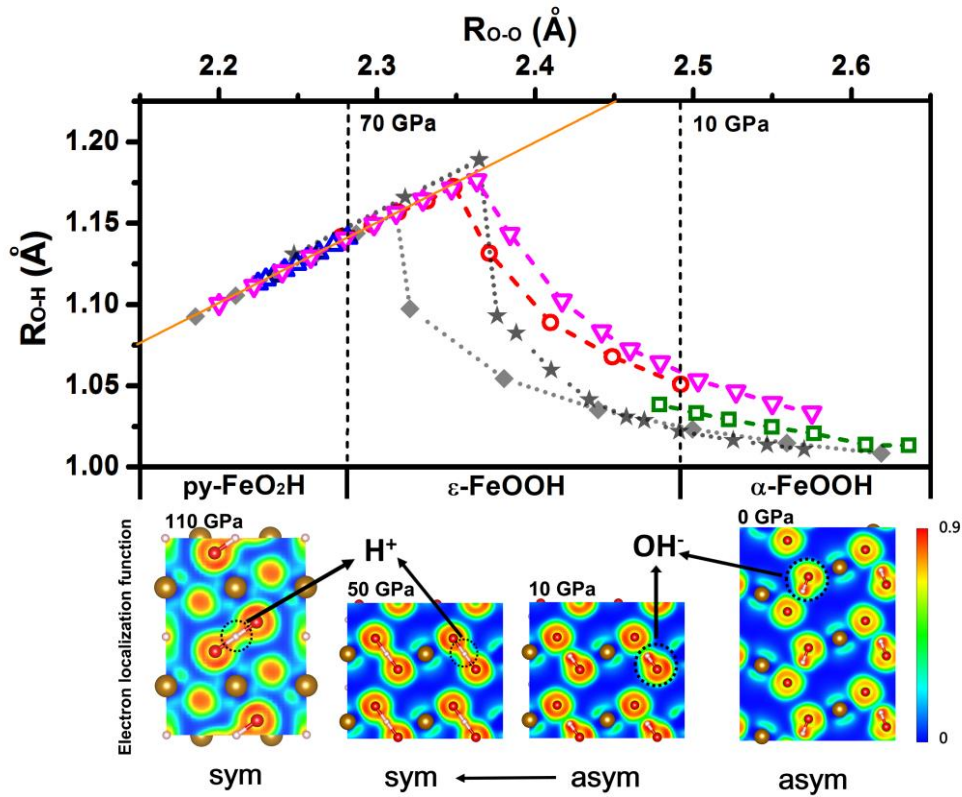

**Figure 2 | Evolution of O-H bond length ($R_{O-H}$) as a function of nearest-neighbor O-O distance ($R_{O-O}$) under compression.** Green squares: α-FeOOH; red circles: ε-FeOOH; blue triangles: py-FeO$_2$H; Magenta inverted triangles: δ-AlOOH; diamonds: ice (VII to X); stars: ice (VIII to X) at 100 K simulated by M. Benoit et al.[23]; orange solid line: line with symmetric O-H-O bonds ($R_{O-H}=R_{O-O}/2$). Dashed lines separate different iron hydrate phases within different pressure ranges: α-FeOOH (Goethite; 0-10 GPa), ε-FeOOH (10-70 GPa) and py-FeO$_2$H (above 70 GPa). The corresponding electron localization function (ELF) distributions viewed along the [001] direction for α-FeOOH, [001] direction for ε-FeOOH, and [111] direction for py-FeO$_2$H are shown below. Dark yellow, white, and red spheres represent Fe, H and O atoms, respectively. The ELF distributions of δ-AlOOH are similar to ε-FeOOH as shown in Extended Data

Fig. 2.

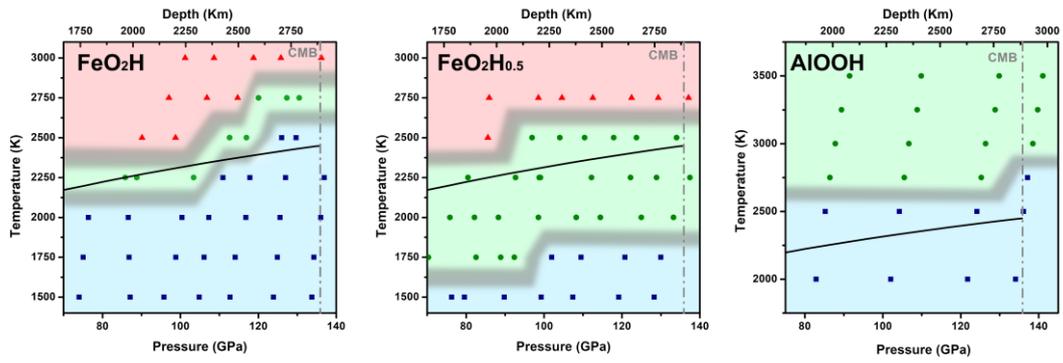

**Figure 3 | Phase diagrams of py-FeO$_2$H, py-FeO$_2$H$_{0.5}$ and δ-AlOOH at temperatures and pressures in the range 1500-3500~K and 70-140 GPa.** The blue, green and red regions indicate the solid, superionic and melting states, respectively. The grey region is uncertainty. The Geotherm for the mantle is shown as black thick solid curves[26].

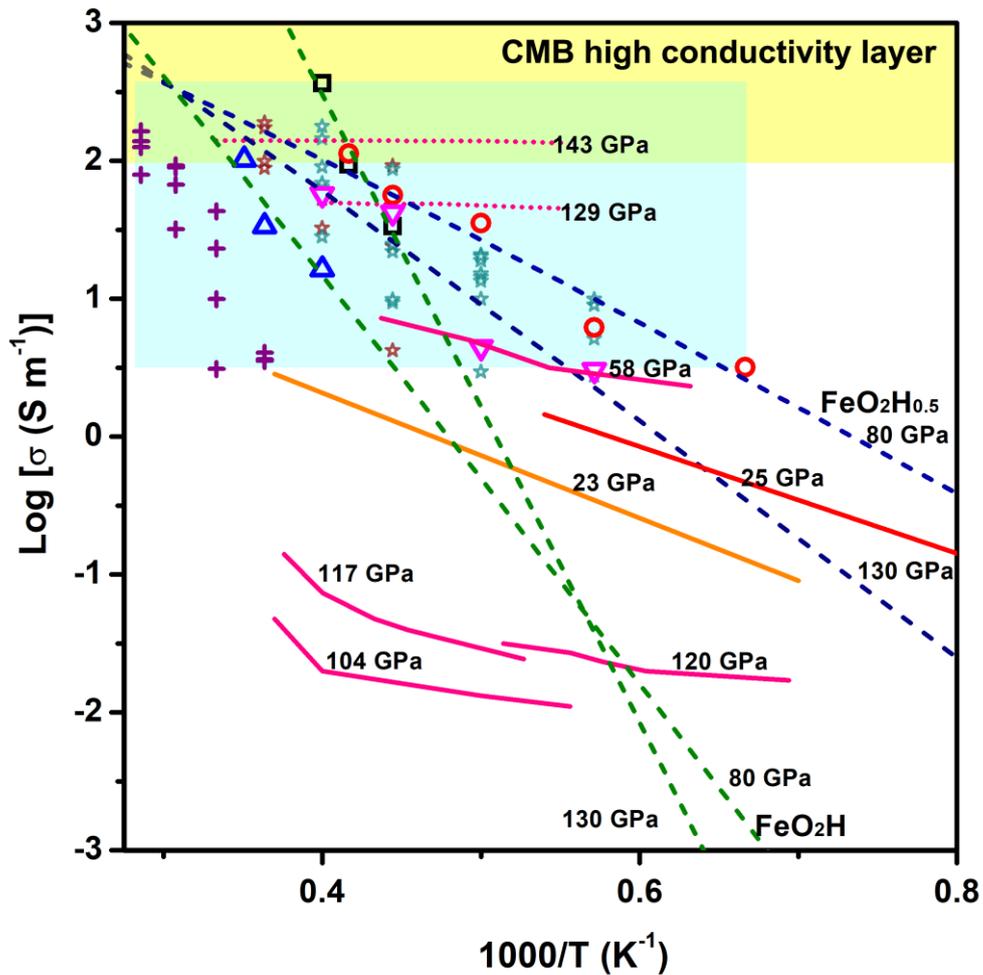

**Figure 4 | Calculated ionic conductivities of py-$FeO_2H_x$ (x = 0.5 and 1) and δ-AlOOH compared with the measured conductivities of Bridgmanite and the post-perovskite phase.** Different conductivities are labeled by different symbols. Black squares: $FeO_2H$ at 80GPa; red circles: $FeO_2H_{0.5}$ at 80GPa; blue triangles: $FeO_2H$ at 130 GPa; Magenta inverted triangles: $FeO_2H_{0.5}$ at 130 GPa; small brown stars: $FeO_2H$ of 7 ps simulations; small cyan stars: $FeO_2H_{0.5}$ of 7 ps simulations; purple crosses: δ-AlOOH; green dashed lines: fitted ionic conductivity of $FeO_2H$ at 80 and 130 GPa; blue dashed lines: fitted ionic conductivity of $FeO_2H_{0.5}$ at 80 and 130 GPa; pink solid lines: bridgmanite $(Mg_{0.9}Fe_{0.1})SiO_3$[27]; pink dotted line: post-perovskite $(Mg_{0.9}Fe_{0.1})SiO_3$[27];

orange solid lines: bridgmanite $(Mg_{0.93}Fe_{0.07})SiO_3$[28]; red solid line: bridgmanite $(Mg_{0.83}Fe_{0.17})SiO_3$[29]. Light blue region shows the conductivity and temperature range of superionic hydrogen. Yellow region shows the conductivity range of high conductivity layer at the CMB.

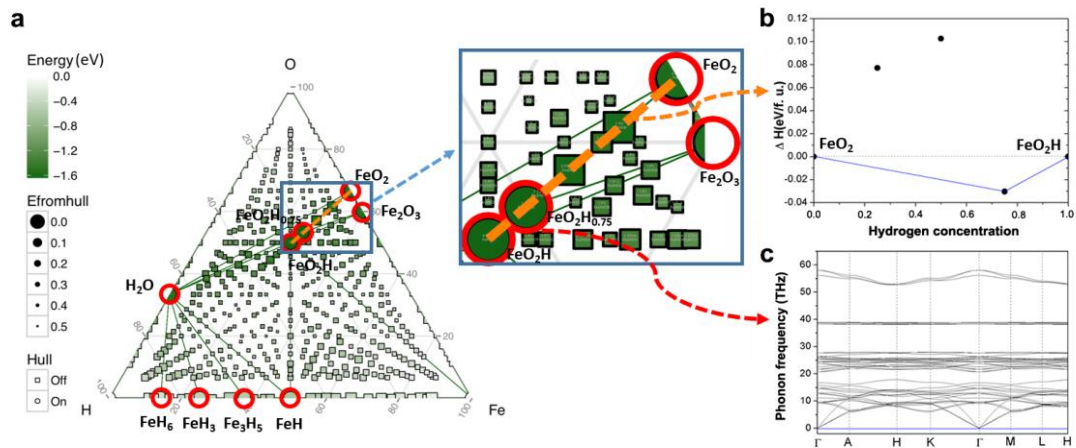

**Extended Data Figure 1 | Convex hull of ternary compounds of Fe, O, and H at 100 GPa. a,** As one can see from the ternary diagram, compounds on the convex hull (ground state phases) are displayed with circles and others are shown in rectangular shaped symbols. The size of rectangles is proportional to the distance from the convex hull surface. The inset on the left top corner shows a magnified segment of convex hull line along the $FeO_2$ and $FeO_2H$. **b,** a partial convex hull along the orange dashed line from $FeO_2$ to $FeO_2H$. **c,** the calculated phonon dispersion curves of $FeO_2H_{0.75}$

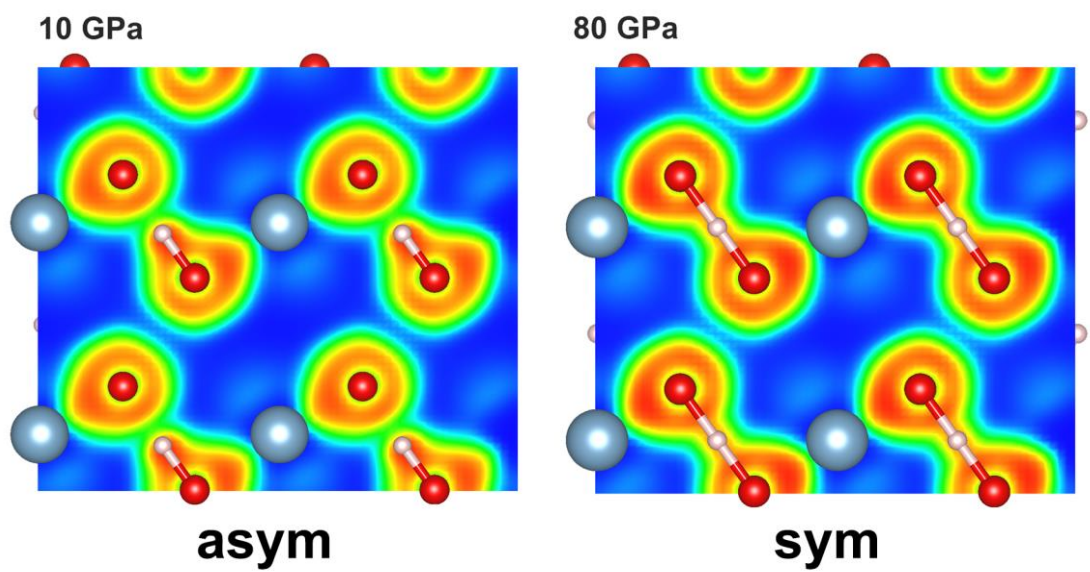

**Extended Data Figure 2 | The electron localization function (ELF) distributions of δ-AlOOH at 10 and 80 GPa.** Viewed along the [001] direction with light blue, white, and red spheres representing Al, H and O atoms, respectively.

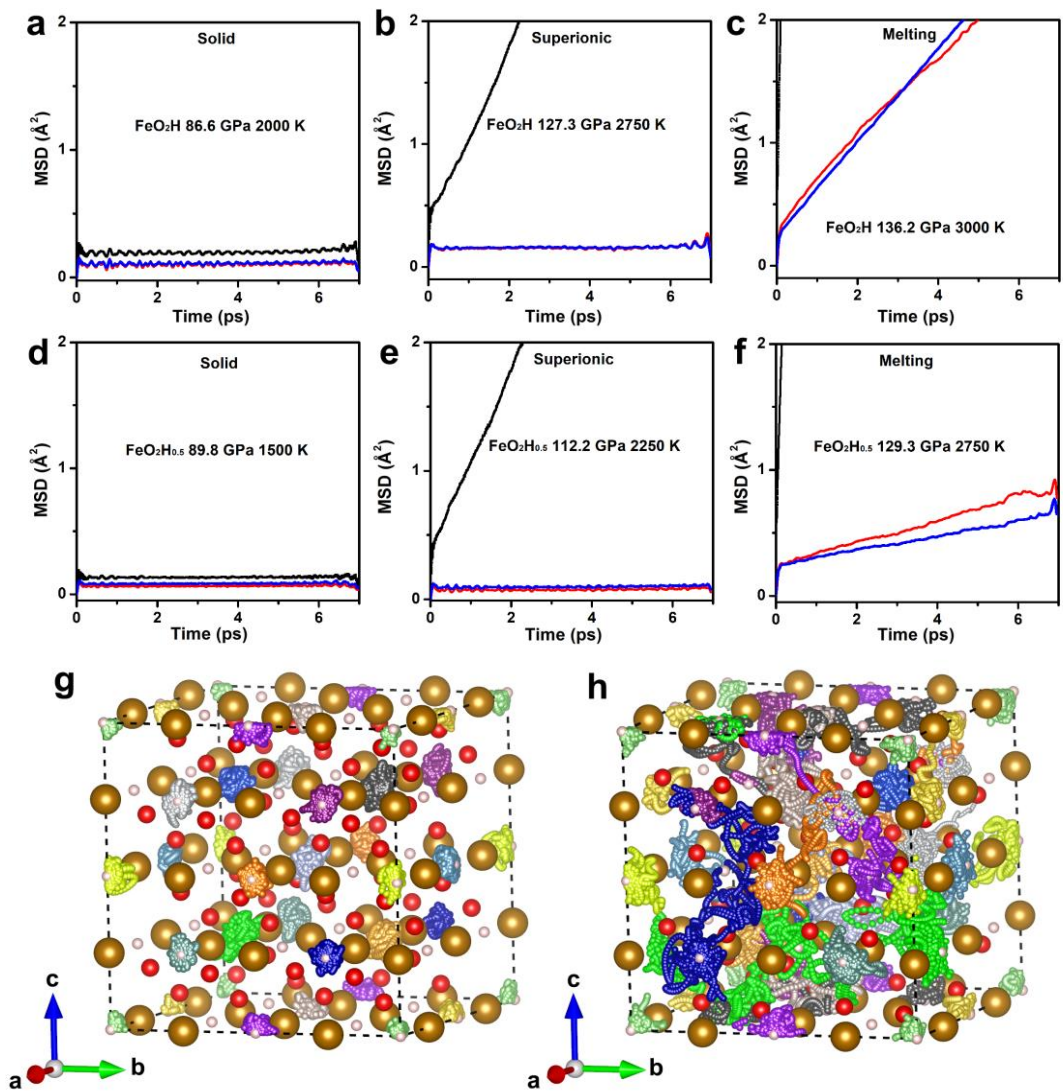

**Extended Data Figure 3 | MSDs and trajectories of protons in py-FeO$_2$H and py-FeO$_2$H$_{0.5}$. a,** FeO$_2$H at 86.6 GPa and 2250 K, **b,** FeO$_2$H at 123.0 GPa and 2750 K, **c,** FeO$_2$H at 116.0 GPa and 3000 K, **d,** FeO$_2$H$_{0.5}$ at 73.7 GPa and 1500 K, **e,** FeO$_2$H$_{0.5}$ at 103.8 GPa and 2250 K, **f,** FeO$_2$H$_{0.5}$ at 122.4 GPa and 2750 K, **g,** proton trajectory in FeO$_2$H$_{0.5}$ in the solid state, **h,** proton trajectory in FeO$_2$H$_{0.5}$ in the superionic state. Small coloured spheres are proton trajectories. The different colours represent the proton at different initial positions in FeO$_2$H$_{0.5}$.

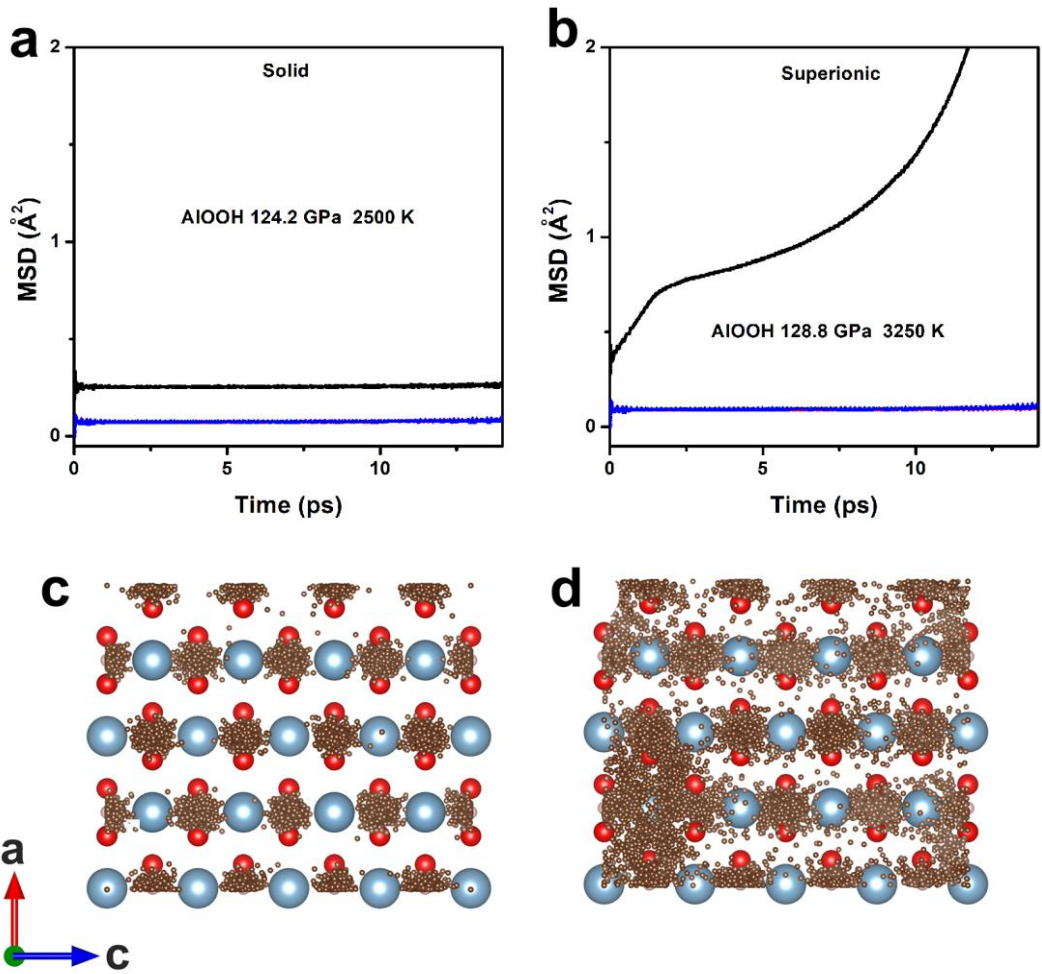

**Extended Data Figure 4 | MSDs and trajectories of protons in δ-AlOOH. a,** 124.2 GPa and 2500 K, **b,** 128.8 GPa and 3250 K, **c,** proton trajectory in δ-AlOOH in the solid state, **d,** proton trajectory in δ-AlOOH in the superionic state. Small brown spheres are proton trajectories.

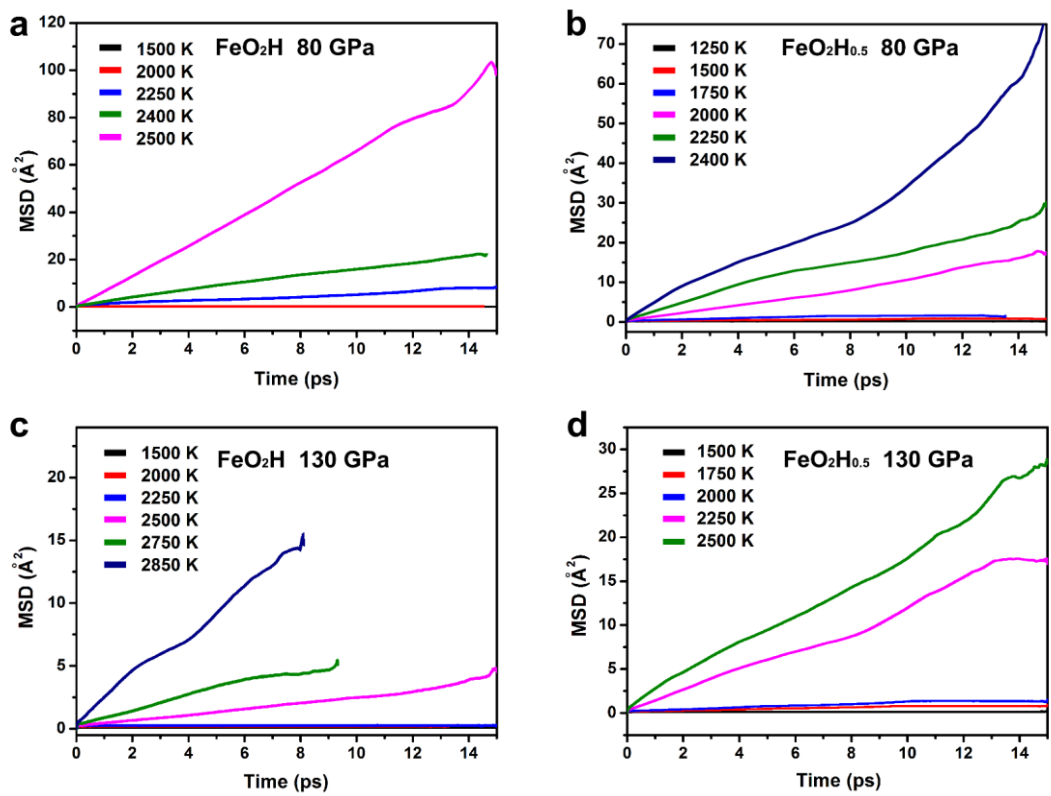

**Extended Data Figure 5 | The MSD of H$^+$ in py-FeO$_2$H$_x$ ( x= 0.5 and 1) at pressures of 80 and 130 GPa and different temperatures. a**, FeO$_2$H at 80 GPa, **b**, FeO$_2$H$_{0.5}$ at 80 GPa, **c**, FeO$_2$H at 80 GPa and **d,** FeO$_2$H$_{0.5}$ at 130 GPa.

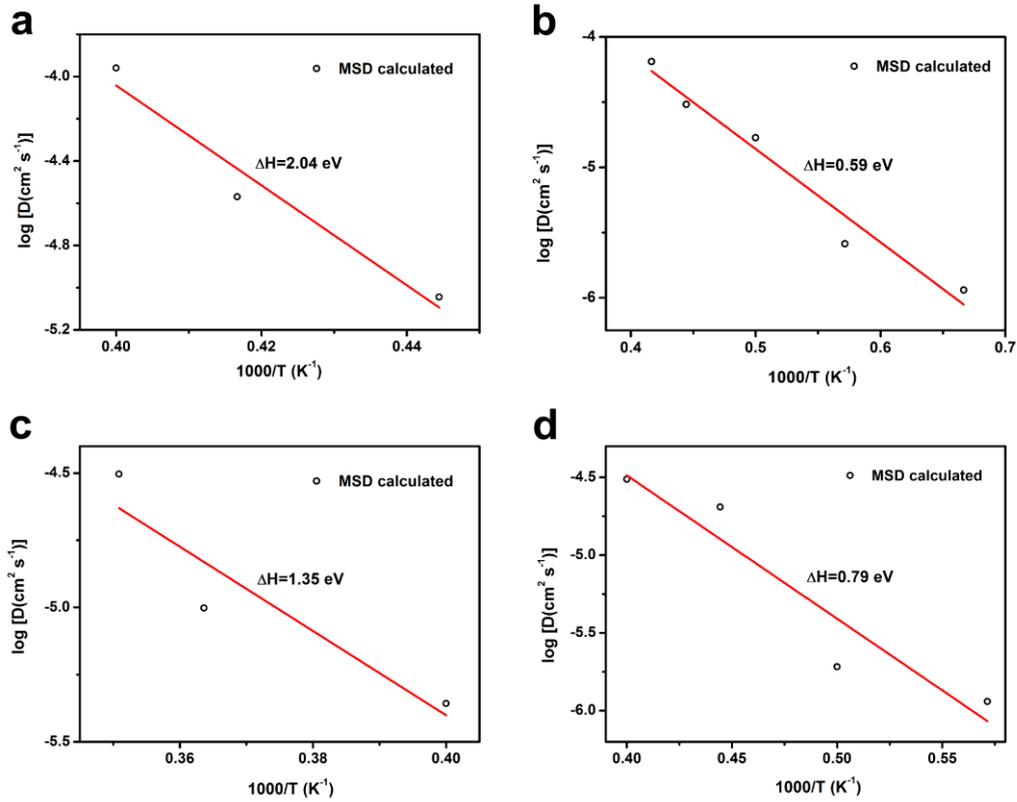

**Extended Data Figure 6 | Diffusion coefficients of proton in py-FeO$_2$H$_x$ (x =1 and 0.5) obtained from MSD.** The data are fitted using Arrhenius equations. The calculated activation enthalpies (ΔH) are given in each image. **a,** FeO$_2$H at 80 GPa, **b,** FeO$_2$H$_{0.5}$ at 80 GPa, **c,** FeO$_2$H at 130 GPa and **d,** FeO$_2$H$_{0.5}$ at 130 GPa.

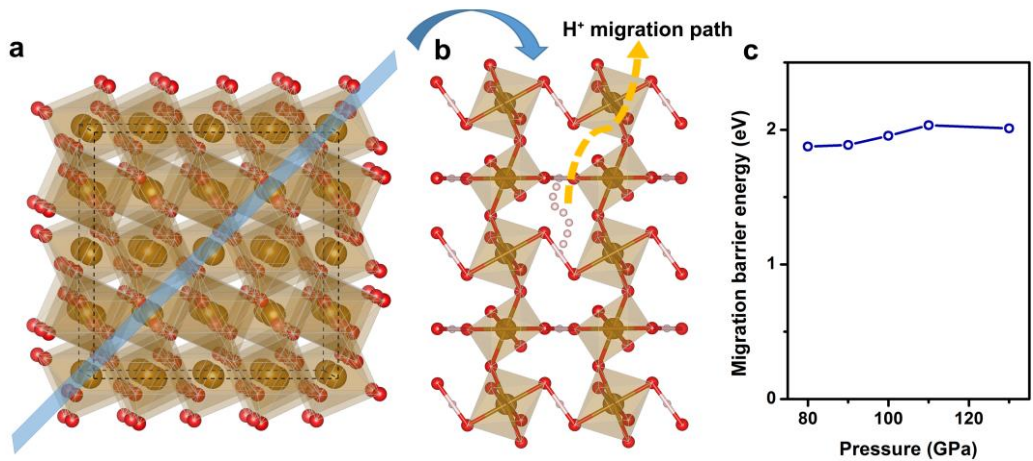

**Extended Data Figure 7 | Proton migration path and barrier energy in the py-FeO₂Hₓ lattice at pressures from 80 to 130 GPa. a,** The crystal structure of $FeO_2H_x$. Brown, red and pink spheres represent Fe, O and H atoms, respectively. **b,** (110) slice cut from Figure 3a and proton hopping of a vacancy along <110> directions, with the minimum energy migration path labeled by a yellow arrow. **c,** the calculated energy barriers of proton migration along the path at pressures of 80, 90, 100, 110 and 130 GPa, respectively.

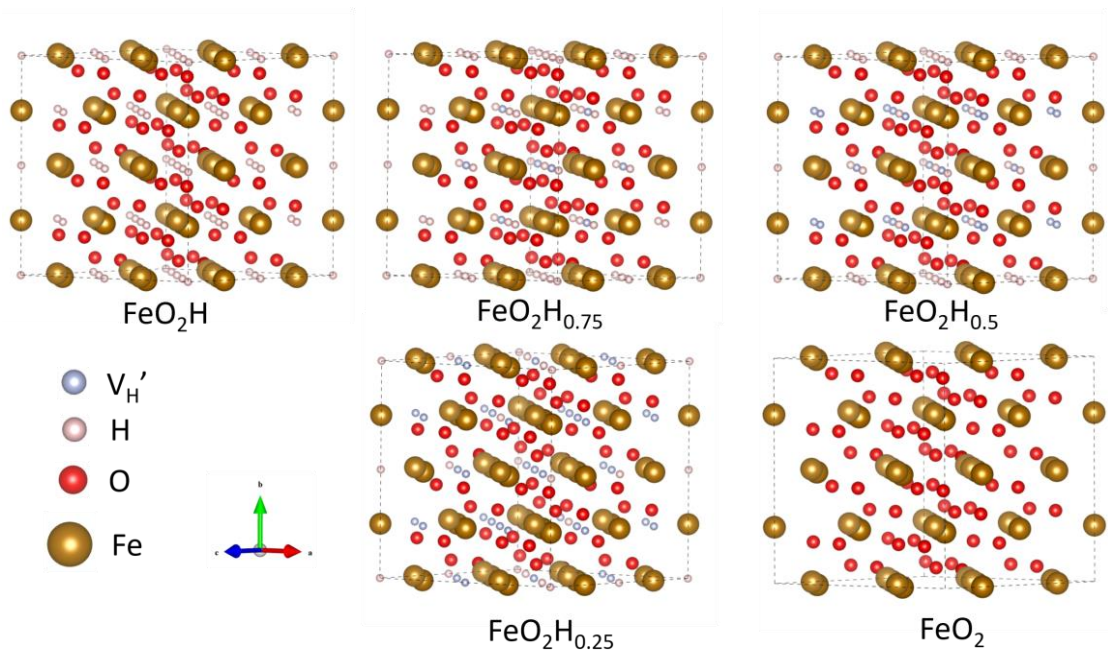

**Extended Data Figure 8 | The crystal structure of py-FeO$_2$H$_x$ (x = 1, 0.75, 0.5, 0.25 and 0).** Light blue, pink, red and brown spheres represent hydrogen vacancies, hydrogen, oxygen and iron atoms.

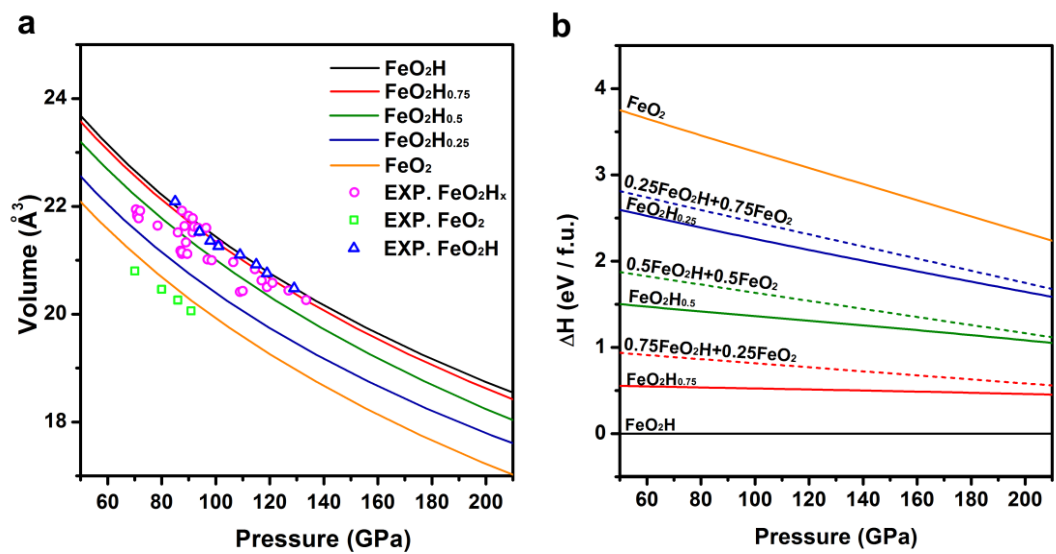

**Extended Data Figure 9 | Calculated P-V relations and relative enthalpies of py-FeO$_2$H$_x$ (x = 1, 0.75, 0.5, 0.25 and 0). a,** the calculated P-V relations (solid lines) are compared with experimental results from previous reports (separated symbols: Magenta circles[10], green squares[19], and blue triangles[9]). **b,** relative enthalpies with respect to py-FeO$_2$H in the pressure range 50-210 GPa.

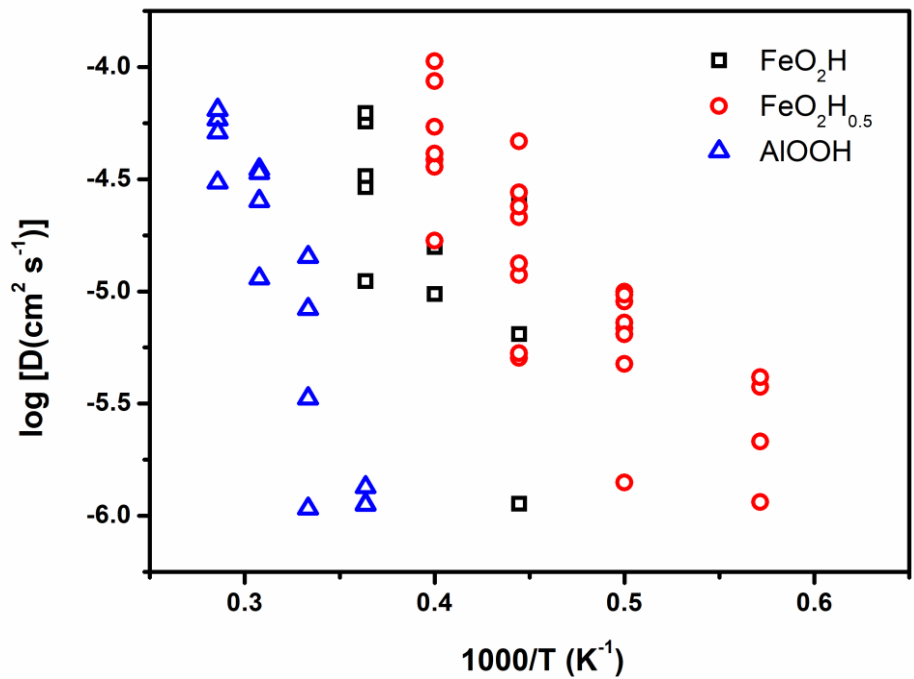

**Extended Data Figure 10 | Diffusion coefficients of proton in py-FeO$_2$H$_x$ (x =1 and 0.5) and δ-AlOOH obtained from MSD.**

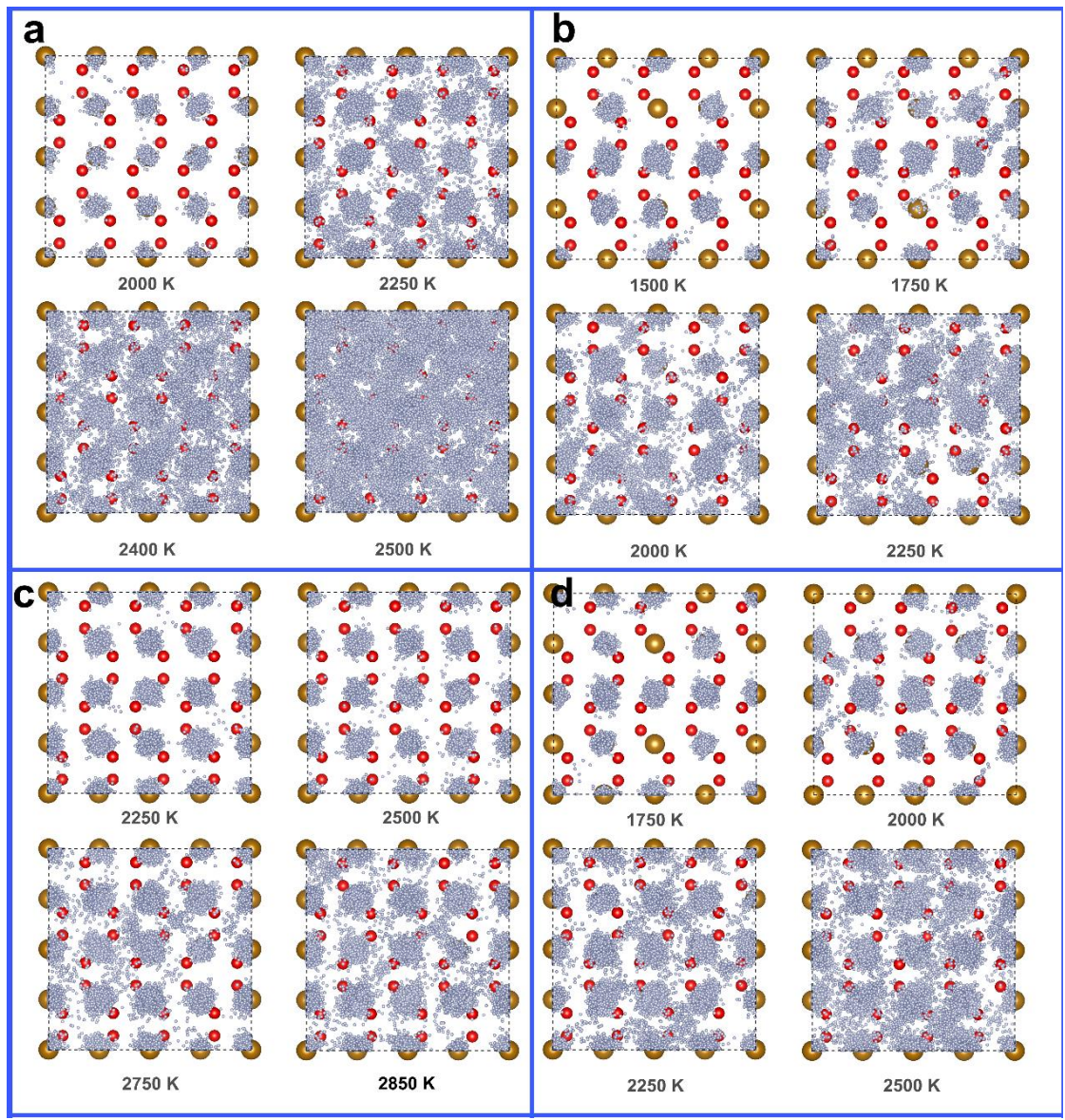

**Extended Data Figure 11 | Trajectories of H$^+$ in py-FeO$_2$H$_x$ (x= 0.5 and 1) at pressures of 80 and 130 GPa for different temperatures.** The small light blue spheres represent proton trajectories. **a,** FeO$_2$H at 80 GPa, **b,** FeO$_2$H$_{0.5}$ at 80 GPa, **c,** FeO$_2$H at 80 GPa and **d,** FeO$_2$H$_{0.5}$ at 130 GPa.

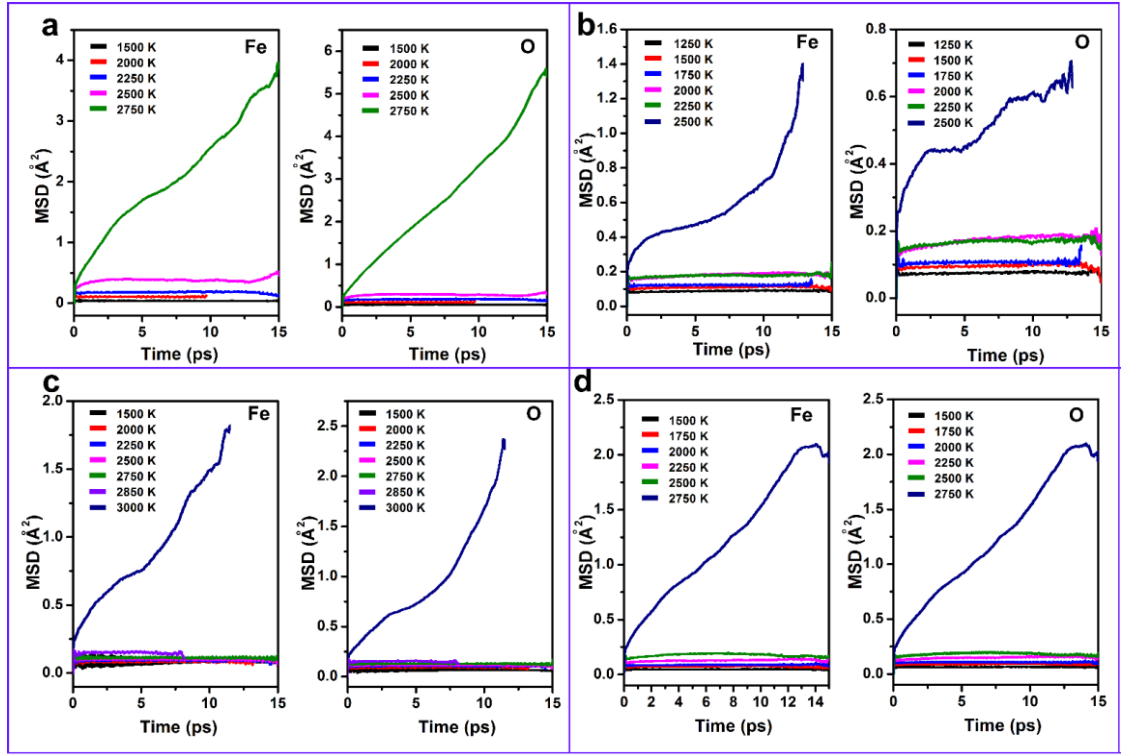

**Extended Data Figure 12 | The MSD of Fe and O ions in py-FeO$_2$H$_x$ (x = 1 and 0.5) at pressures of 80 and 130 GPa and different temperatures.** The obvious increase in the MSD of Fe and O ions indicates melting. **a,** FeO$_2$H at 80 GPa, **b,** FeO$_2$H$_{0.5}$ at 80 GPa, **c,** FeO$_2$H at 130 GPa and **d,** FeO$_2$H$_{0.5}$ at 130 GPa.

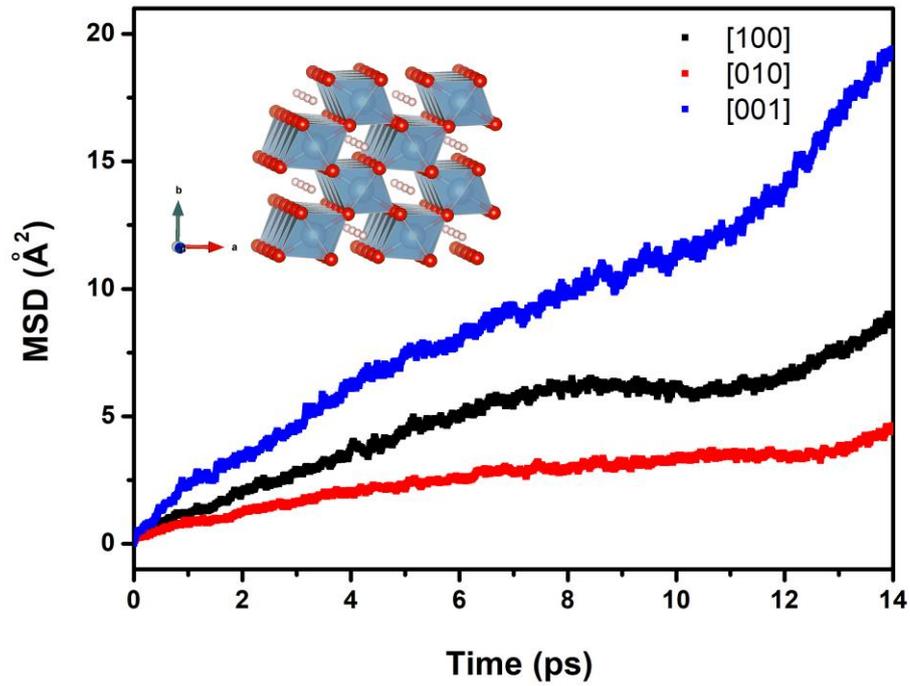

**Extended Data Figure 13 | The MSD of superionic protons in δ-AlOOH along [100], [010] and [001] directions.** Anisotropy superionic proton diffusion in δ-AlOOH ascribing to the channel-like lattice structure of δ-AlOOH as shown in the inset (Light blue, white, red spheres represent Al, H and O atoms).

**Extended Data Table 1 | Comparison of calculated bulk modulus, the first pressure derivative of bulk modulus, equilibrium volume and equilibrium total energy of py-FeO$_2$H$_x$ (x = 1, 0.75, 0.5, 0.25, 0) and δ-AlOOH.**

| Phase | B$_0$ (GPa) | B$_0$' | V$_0$ (Å$^3$/f.u.) | E$_0$ (eV/f.u.) |
|---|---|---|---|---|
| FeO$_2$H | 232.7 | 4.01 | 27.70 | -21.45 |
| FeO$_2$H$_{0.75}$ | 248.2 | 3.81 | 27.41 | -20.84 |
| FeO$_2$H$_{0.5}$ | 270.33 | 3.50 | 26.79 | -19.75 |
| FeO$_2$H$_{0.25}$ | 232.82 | 3.94 | 26.40 | -18.48 |
| FeO$_2$ | 257.46 | 3.45 | 25.66 | -17.15 |
| AlOOH | 211.02 | 4.00 | 14.20 | -13.04 |